\documentclass{aastex62}

\submitjournal{ApJ}
\usepackage{color}

\begin{document}

\title{An Alternative Channel For High-Mass Binary Black Holes-----Dark Matter Accretion Onto Black Hole}

\correspondingauthor{Guoliang L\"{u}}\email{guolianglv@xao.ac.ac}

\author{Tongzheng Wang}
\affil{School of Physical Science and Technology,
Xinjiang University, Urumqi, 830046, China}

\author{Lin Li}
\affil{School of Physical Science and Technology,
Xinjiang University, Urumqi, 830046, China}

\author{Chunhua Zhu}
\affil{School of Physical Science and Technology,
Xinjiang University, Urumqi, 830046, China}

\author{Zhaojun Wang}
\affil{School of Physical Science and Technology,
Xinjiang University, Urumqi, 830046, China}

\author{Anzhong Wang}
\affil{GCAP-CASPER, Physics Department, Baylor University, Waco, Texas 76798-7316, USA}
\affil{Institute for Advanced Physics and Mathematics, Zhejiang University of Technology, Hangzhou, 310032, China}

\author{Qiang Wu}
\affil{Institute for Advanced Physics and Mathematics, Zhejiang University of Technology, Hangzhou, 310032, China}

\author{Hai-Shan Liu}
\affil{Institute for Advanced Physics and Mathematics, Zhejiang University of Technology, Hangzhou, 310032, China}

\author{Guoliang L\"{u}}
\affil{School of Physical Science and Technology,
Xinjiang University, Urumqi, 830046, China}



\begin{abstract}
By a method of population synthesis we construct a model of dark matter (DM) accretion onto binary black holes (BHs),
and investigate the merger rate of the binary BHs.
We find that the merger rate can weakly increase (less than 10\%).
However, the DM accretion can efficiently enhance the masses of binary BHs.
In our model, the result for $Z = 0.01$ without the DM accretion can not explain the GW170104, GW170814, and GW150914,
while with the DM accretion, it can cover all observations well.
For the higher metallicity ($Z = 0.02$),
our model can not explain the mergers of high-mass binary BHs like GW170104, GW170814, and GW150914.
We estimate that the merger rate of binary BHs lies between $\rm55Gpc^{-3}yr^{-1}$ to $\rm197Gpc^{-3}yr^{-1}$.
\end{abstract}



\section{INTRODUCTION}

On 2015 September 14 09:50:45, the advanced Laser Interferometer Gravitational-Wave Observatory (aLIGO)
detected its first gravitational wave event, which was named as GW150914 \citep{Abbott2016a}. This
transient signal was produced by the merger of two BHs. Their masses are
$M_{1}=36^{+5}_{-4}{\rm M_{\odot}}$ and $M_{2}=29^{+4}_{-4}{\rm M_{\odot}}$, respectively.
The aLIGO have opens a new era for observing the Universe. Simultaneously,
it also challenges popular theoretical scenarios, especially, the theory of stellar evolution.
Not only from the point of view of theoretical simulations but also observational estimates, it is very difficult
to produce so heavy BHs. Theoretically, most of BHs have masses lower than 10$\rm M_\odot$ unless
the stellar metallicity is very low \citep[e. g.,][]{Zhang2008,Fryer2012}.
\cite{Ozel2010} examined 16 low-mass X-ray binary systems containing BHs. They found that the masses of BHs
hardly exceeded $20\rm M_\odot$ and there was a strongly peaked distribution at $7.8\pm1.2\rm M_{\odot}$.
\cite{Farr2011} obtained similar results.

Soon, aLIGO found another three gravitational wave events (GW151226, GW170104, and GW170814) \citep{Abbott2016b,Abbott2017,Abbott2017b}.
The masses of BHs in GW151226 were $14.2^{+8.3}_{-3.7}\rm M_{\odot}$ and
$7.5^{+2.3}_{-2.3}\rm M_{\odot}$, respectively \citep{Abbott2016b}.
They can easily be explained by popular theory of stellar evolution \citep{Stevenson2017}.
However, there were two heavy BHs in GW170104 and GW170814, and their masses were ($31.2^{+8.4}_{-6.0}\rm M_{\odot}$, $19.4^{+5.3}_{-5.9}\rm M_{\odot}$),
and ($30.5^{+5.7}_{-3.0}\rm M_{\odot}$, $25.3^{+2.8}_{-4.2}\rm M_{\odot}$), respectively \citep{Abbott2017, Abbott2017b}.

In order to produce these heavy BHs ($M_{\rm BH}>\sim20\rm M_\odot$), several evolutional scenarios have been put forward. They include:
the binary evolution channel with failed supernova model \citep{Belczynski2016,Spera2017}
and with chemical homogeneous evolution \citep{Mandel2016,Marchant2016,Eldridge2016},
the dynamical coalescing of BHs in globular, young stellar, or nuclear star clusters \citep{Rodriguez2016,Mapelli2016,Askar2017,Petrovich2017,
Banerjee2017,Rodriguez2017,Antonini2016,Hoang2017,O'Leary2016,Bartos2017,Stone2017},
and the merger of primordial BHs \citep{Sasaki2016,Bird2016}.

The same requirement for heavy BHs originating from stars is the low metallicity which leads to low mass-loss rate or effective
chemical homogeneous evolution \citep[e. g.,][]{Abbott2016b}. The similar goal is to obtain heavy BHs by enhancing the mass of helium core.
Although there are some observational evidences that may support the failed supernova model
or chemical homogeneous evolution \citep{Adams2017,Evans2005,Hunter2009},
the origin of heavy BH still open. Even, we can not rule out GW150914, GW170104, and GW170814 coming from the merger of primordial BHs.

All authors of the above mentioned literatures neglected the increase of BH's mass via accretion matter.
It is easily understood because the binary BHs have merged before significant
mass accretion \citep[e. g.,][]{Tagawa2016}. However, since \cite{Spergel2000} suggested that cold dark matter (DM)
are self-interacting particles. The angular momentum can transport outwards rapidly enough due to self-interacting of dark matter, and super massive BHs ($10^{6}\leq M/\rm M_{\odot}\leq 10^{9} $) can form by the seeds accrete DM \citep[e. g.,][]{Ostriker2000,Hennawi2002,Balberg2002}.
Then, could the heavy BHs originate from normal stellar BHs ($M_{\rm BH}<\sim10\rm M_\odot$) via accreting DM?
There is a great of computational work for resolving the structure of halos in order to compare with
the observational data \citep[e. g.,][]{Navarro1996,Navarro1997,Moore1998,Moore1999}.
Therefore, it is possible for us to discuss an alternative channel for producing heavy BHs -----dark matter accretion.

In this work, we focus on the binary BHs accreting DM because almost all of heavy BHs are discovered in the merger of binary BHs.
In Section 2, we present our assumptions and describe some details of the modelling algorithm
(including stellar evolution, DM accretion onto binary BHs and the method of population synthesis).
The merger rates of binary BHs are given in Section 3. The conclusions appear in Section 4.

\section{Model}
In the model of DM accretion onto binary BHs, the mass of nascent BH,
the distribution structure of DM and the accretion model are critical.

\subsection{Mass of nascent BH }
In the popular theory, the mass of nascent BH depends on the the final CO core mass ($M_{\rm CO}$) which determines FeNi core mass
\citep[e. g.,][]{Hurley2002,Belczynski2008}.
However, $M_{\rm CO}$ depends on some uncertain parameters: stellar mass-loss rate, rotation, and so on.

In our work, we use the Modules for Experiments in Stellar Astrophysics code
(MESA, see \cite{Paxton2011,Paxton2013,Paxton2015} for details.) to simulate stellar evolution.
The following input parameters are used: The Ledoux criterion
is used for convection, mixing-length parameter is taken as 1.5, an efficiency parameter of unity
is assumed for semi-convection. The metallicity ($Z$) greatly affects the mass of nascent BH.
In order to discuss its effects, $Z$ is taken as 0.0001, 0.001, 0.01, and 0.02, respectively.
Following \cite{Zhu2017}, we use the model of \cite{Vink2001} to calculate the mass-loss rates.
Simultaneously, rotation can enhance the mass-loss rate \citep{Langer1998} and induce instability of various kinds
so that trigger chemical homogeneous evolution \citep[e. g.,][]{Heger2000}.
We use similar input parameters with \cite{Zhu2017} (See details in Section 2 of \cite{Zhu2017}).

\begin{figure}
\includegraphics[totalheight=3.5in,width=3.in,angle=-90]{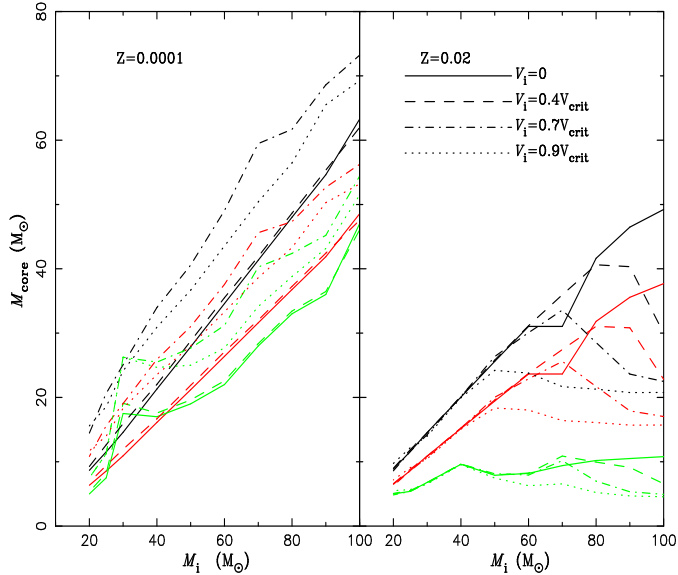}
\caption{Core masses vs initial masses with different metallicities and rotational velocity.
Black, red, and green lines represent the He-core, CO-core, and BH masses, respectively.
BH masses are calculated by the formulae in \cite{Belczynski2008}.
Solid, dashed, dot-dashed, and dot lines mean that the rotational velocity at zero-age main sequence
equal 0, 0.4$V_{\rm crit}$, 0.7$V_{\rm crit}$, and 0.9$V_{\rm crit}$, respectively.
Here, $V_{\rm crit}$ is the critical rotational velocity. }
\label{fig:corm}
\end{figure}

We use MESA to calculate the He-core and CO core masses of stars with different
$Z$s ($Z=0.0001$ and $Z=0.02$), initial masses ($M_{\rm i}=20, 25, 30, 40, 50, 60, 70, 80, 90,$ and 100 $M_\odot$)
and velocities ($V_{\rm i}=0$, 0.4$V_{\rm crit}$, 0.7$V_{\rm crit}$, and  0.9$V_{\rm crit}$.
Here $V_{\rm crit}$ is the critical rotational velocity),
which is showed in Figure \ref{fig:corm}.
Obviously, both of the metallicity and the rotational velocity affect the stellar He-core and CO-core masses.
Especially, the rotational velocity have very different effects for different metallicities.
For low metallicity ($Z=0.0001$), from $V_{\rm i}=0$ to $V_{\rm i}=0.7V_{\rm crit}$,
the He-core masses increases by about 50\%. The main reason is that the high rotational velocity
induces more efficient the chemical homogeneous evolution which
results in the expansion of helium produced by nuclear reactions.
However, higher rotational velocity ($V_{\rm i}=0.9V_{\rm crit}$) enhances too high
mass-loss rate. Compared with the model of $V_{\rm i}=0.7V_{\rm crit}$, the increase of He-core
the mass decreases.
For high metallicity ($Z=0.02$), high rotational velocity can not produce
efficiently chemical homogeneous evolution but high mass-loss rate \citep{Heger2000,Mandel2016,Marchant2016}.

Considering that there are many uncertainties and there always are some convergent problems to
simulate core-collapse supernova, we use CO core masses to calculate the BH masses by the
formulae in \cite[]{Belczynski2008} [See Eqs. (1) and (2)]. The green lines in Figure \ref{fig:corm}
give the BH masses for different initial-mass stars with different rotational velocities.

Although MESA can simulate the evolution of some binary systems, it can not deal with some dynamical
processes, for example, common envelope episodes. Even it breaks down when calculate
binary evolution with high mass-transfer rate ($>\sim 10^{-3}\rm M_\odot $yr$^{-1}$).
Therefore, in our work, we use rapid binary evolution (BSE) code, originating
from \cite{Hurley2002}, to simulate binary evolution with mass transfer.
For a given binary system, including the primary mass ($M_1$)
and its initial rotational velocity $V_1$, the secondary mass ($M_2$) and its initial rotational velocity $V_2$,
the binary separation ($a$), we use MESA to simulate its evolution if the binary is always detached.
If any companion in binary system fills its Roche lobe, we use BSE to simulate mass-transfer evolution.
When the binary become detached again, BSE can give all binary parameters. After that, we use MESA to
simulate its evolution.

In binary model, there are many input parameters which can affect the binary evolution.
In these parameters, the kick velocity, added to a nascent neutron star or BH during core-collapse
supernova, has great effects on the formation of binary BHs.
Based on the proper motion of 233 pulsars, \cite{Hobbs2005} considered that the kick velocity
has a Maxwellian distribution with $\sigma_{\rm k}=265$ km s$^{-1}$.
Using the kick velocity distribution of \cite{Hobbs2005}, we can calculate the kick velocity ($v_{\rm k}$) of a nascent neutron star or BH.
However, there are some growing evidences that the kick velocities of BHs are smaller than these of neutron stars \citep{Mandel2016b}.
Following \cite{Eldridge2016}, we assume that the true kick velocity of a nascent BH ($v_{\rm k}^{BH}$) equals $v_{\rm k}$ $(\frac{1.4}{M_{\rm BH}})$,
where $M_{\rm BH}$ is the mass of nascent BH in solar unit.

\subsection{DM accretion onto binary BHs}
Combining MESA and BSE code, we can get binary BHs, including masses of two BHs, and orbital period and eccentricity.
In our model, masses of two BHs increase via accreting DM, and the orbital period shrinks due to gravitational release.
In order to calculate the DM accretion, we must know the density structure of DM.

Using high-resolution N-body simulations, \cite{Navarro1997} gave the density structure
of DM in hierarchically clustering universes.
It can be well described by a numerical simulation with two free parameters given by
\begin{equation}
  \rho_{\rm D}(r)=\rho_{\rm D\odot}\frac{(\rm D_{\odot}/\it r_{s})[1+(\rm D_{\odot}/\it r_{s})]^{2}}{(r/r_{s})[1+(r/r_{s})]^{2}},
  \label{dm}
\end{equation}
where $\rm\rho_{D\odot}$ is the local dark matter, $\rm D_{\odot}$ is the distance of
the Sun away from the Galactic center and $r_{s}$ is the scaling radius.
Following \cite{Bernal2012}, we take $\rm\rho_{D\odot}=0.4Gev/cm^{3}$, $\rm D_{\odot}=8.3kpc$, and $r_{s}=20\rm kpc$.

Although the nature of DM is still unclear, it unquestionably produces gravitational force. And as we all know Bondi accretion is suitable for spherically symmetric accretion onto a star, but for the accretion of BHs bind in a binary is not complete spherically symmetric. It is beyond the scope of this work to conceive how to solve the problem. In this paper, we treat DM accretion as quasi-spherical accretion. Using the model of Bondi accretion to estimate the DM accretion-rate by BH via \cite{Bondi1952}
 \begin{equation}
  \dot{M}=4\pi\frac{G^{2}M^{2}}{C^{3}_{\rm A}}\rho_{\rm D},
  \label{Accr}
\end{equation}
where $\rho_{\rm D}$, $C_{A}$ are the density and sound speed in the surrounding of the DM fluid, respectively.
$M$ is the mass of BH, $\rm G$ is the constant of gravitation.

We can solve Eq. \ref{Accr} to obtain
 \begin{equation}
  M_{t}=\frac{M_{0}}{1-t/\tau}
  \label{mt}
\end{equation}
where

 \begin{equation}
  \tau=\frac{C_{\rm A}^{3}}{4 \pi \rm G^{2} \it \rho_{\rm D} M_{0}}
  \label{t}
\end{equation}

$M_{0}$ is the initial mass of the black hole, $M_{t}$ is the the mass of the black hole after dark matter accretion. Following \cite{Ostriker2000}, we take
 $C_{\rm A}=100\rm km/s$.
If $M_{0}=15 \rm M_{\odot}$ and $\rho_{\rm D}=15000 \rm M_{\odot}/pc^{3}$,
then one get $\tau=1.7\times10^{10} \rm yr$.

We can also present this expression as a function of $r$
 \begin{equation}
  \tau=r(\rm pc){[1+\frac{\it r \rm(pc)}{20000(pc)}]\times1.6\times10^{3}\rm} Gyr
  \label{rt}
\end{equation}
Using function \ref{mt} and function \ref{rt}, one can estimate the binary BHs mass at mergers with the Galactic plane when the initial mass of black hole is $15 \rm M_{\odot}$.
\begin{figure}
\includegraphics[totalheight=3.4in,width=3.4in,angle=-90]{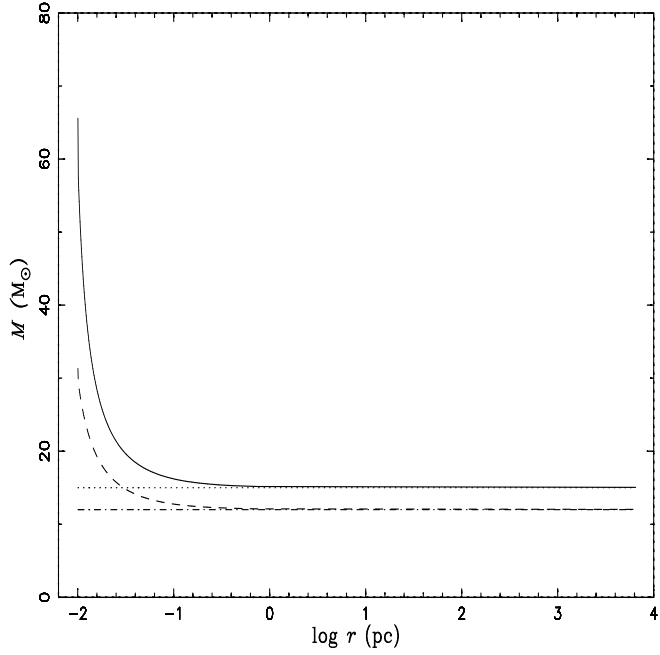}
\caption{
The black hole mass with dark matter accretion along the Galactic plane.
The dot-dished and dotted lines show the initial mass of black hole($M_{0}=12\rm M_{\odot}$ and $M_{0}=15\rm M_{\odot}$). The dashed and solid lines show the mass of black hole after dark matter accretion.}
\label{fig2}
\end{figure}

The DM accretion can increase BH masses, and also may change the orbital angular momentum ($J_{\rm orb}$) of binary BHs.
However, we do not know the nature of DM. Therefore, we assume that the DM accretion do not change
$J_{\rm orb}$, and its decay is only determined by gravitational release.
The decay ratio of $J_{\rm orb}$ was given by \cite{Faulkner1971} via
\begin{equation}
  \frac{\dot{J_{\rm GB}}}{J_{\rm orb}}=-\frac{32 \rm G^{3}}{5\rm c^{5}}\frac{M_{1}M_{2}M}{a^{4}}
  \label{gr}
\end{equation}
where $\rm c$ is speed of light, $ a$ is the separation of binary stars
and $M$ is the total mass of two BHs.
We assume that two BHs merge when their separation is less than the sum of their Schwarzschild radius,
and the Schwarzschild radius of BH is given by
\begin{equation}\label{}
  r_{\rm BH}=\frac{2{\rm G} M}{\rm c^{2}}.
\end{equation}

\begin{figure}
\includegraphics[totalheight=2.4in,width=3.4in,angle=0]{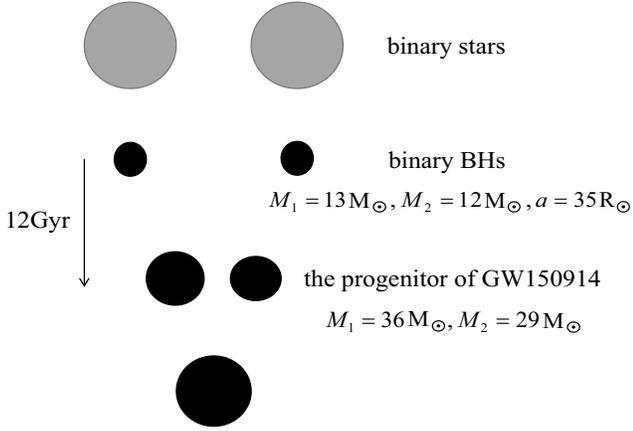}
\caption{Example of binary BHs merger similar to GW150914.
The mass of nascent binary BHs are $\rm13M_{\odot}$ and $\rm12M_{\odot}$, and their separation is $\rm35 R_{\odot}$, then the binary BHs accrete DM and merge by the gravitational wave radiation.
After 12Gyr evolution, the binary BHs mass are $\rm36M_{\odot}$ and $\rm29M_{\odot}$.}\label{yh}
\end{figure}

Figure \ref{yh} shows an example for binary BHs accreting DM via Eq. (\ref{Accr}) and merging through the release of
gravitational waves via Eq. (\ref{gr}).
The masses of nascent binary BHs are $\rm13M_{\odot}$ and $\rm12M_{\odot}$, respectively,
and their separation is $\rm 35 R_{\odot}$.
If we place this binary BHs in a position away 0.01 pc from the Galactic center,
where $\rho_{\rm D}=\rm18000M_{\odot}/pc^{3}$, the masses of the binary BHs increases to $\sim 36\rm M_{\odot}$ and $\sim 29\rm M_{\odot}$,
respectively. At the same time, the binary BHs begin to merge after 12.0 Gyr with the gravitational release.
This system of binary BHs should be a possible progenitor of GW150914.

\section{Merge Rates of Binary BHs.}

In order to investigate the merge rate of binary BHs,
we carry out binary population synthesis via a Monte Carlo simulation technique.
Similar to the main case considered in a series of
our studies \citep{Lu2006,Lu2008,Lu2009,Lu2012,Lu2013}, we use
the initial mass-function of \citep{Miller1979} for the mass
of the primary components and a flat distribution of mass ratios
\citep{Mazeh1992,Goldberg1994}. The distribution of
separations is determined by $\log a = 5X + 1$, where $X$ is a random
variable uniformly distributed in the range [0, 1] and the separation
$a$ is in $\rm R_\odot$.

First, we estimate the merge rates of binary BHs in the Milk Way.
A constant star-formation rate (SFR) of  $4.0 \rm M_{\odot}{\rm yr}^{-1}$ for 13 Gyr in
the Milk Way is assumed. The distribution of these stars in the Milk Way
follows the flattened density double power-law model given by \cite{Zhao1997}
\begin{equation}\label{}
  \rho(r)=A^{*}(r/r_{\rm core})^{-\gamma}[1+(r/r_{\rm core})^{\alpha}]^{(\gamma-\beta)/\alpha},
  \label{eq:nrho}
\end{equation}
where
\begin{equation}\label{}
 A^{*}=\rho_{\odot}(r_{\odot}/r_{\rm core})^{\gamma}[1+(r_{\odot}/r_{\rm core})^{\alpha}]^{(\beta-\gamma)/\alpha}
\end{equation}
and $r^{2}=X^{2}+(Y/p)^{2}+(Z/q)^{2}$. The $r$ is an axisymmetric radius, and $(X, Y, Z)$ are the Galactic centric cartesian coordinates.
In Eq. (\ref{eq:nrho}), there are six free parameters: $\alpha, \beta, \gamma, r_{\rm core}, q,$ and $\rm\rho_{\odot}$.
The parameters $r_{\rm core}$ and $\rm\rho_{\odot}$ are a scaling radius and the local density of the stellar halo
in the solar neighborhood, respectively. The parameters $p$ and $q$ are the axis ratios.
Based on the survey of 2MASS and SDSS-III/APOGEE, \cite{Fern2015} estimated the magnitudes of the above six parameters:
$\alpha=1$, $\beta=3.76$, $\gamma=1$, $ r_{\rm core}=2180\rm pc$, $\rm\rho_{\odot}=4.14\times10^{-5}M_{\odot}/pc^{3}$,
$p=1$, and $q=0.77$.
Considering the DM profiles being independent of halo mass,
we take Eq. (\ref{dm}) as the DM profiles of the Milk Way.

The luminosity distances of all gravitational wave events observed is about 500 Mpc or even more far away from
the Earth. In order to estimate the merger rate of binary BHs beyond the Milk Way,
we use the SFR given by \cite{Strolger2004}
\begin{equation}\label{}
  SFR=10^{9}a(t^{b}e^{-t/c}+de^{d(t-\rm t_{0})/c})\rm M_{\odot}yr^{-1}Gpc^{-3}
\end{equation}
where $t$ is the age of Universe, which is given in Gyr, and $\rm t_{0}$ is the present age of the Universe.
Following \cite{Strolger2004}, we take $\rm t_{0}=13.47Gyr$ and the parameters $a=0.021, b=2.12, c=1.69$, and $d=0.207$.

\begin{figure}
\includegraphics[totalheight=3.2in,width=3.2in,angle=-90]{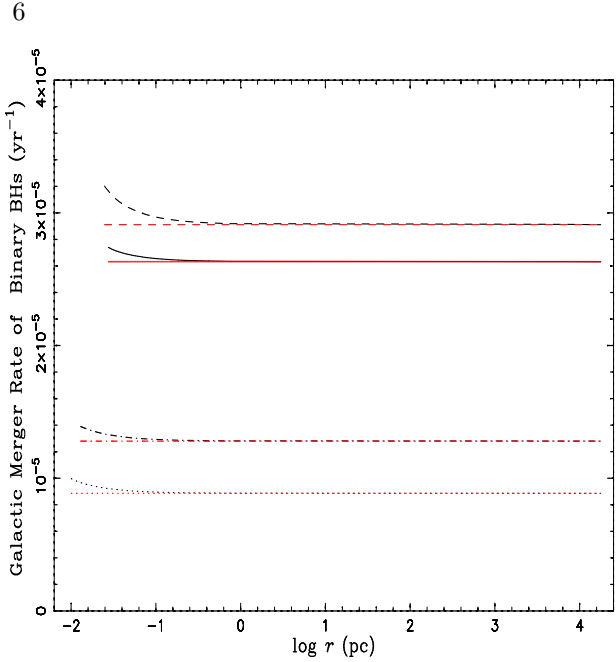}
\caption{Merger rate of binary BHs without (red line) and with (black line) DM accretion along the Galactic plane.
The solid, dished, dish-dot, and dot lines show the merger rate
of $Z=0.0001, 0.001, 0.01$, and 0.02, respectively.}\label{line}
\end{figure}

We calculate the evolution of $5\times10^9$ binary systems for $Z=0.0001, 0.001, 0.01,$ and 0.02, respectively.
The total stellar mass is about $5\times10^9 \rm M_\odot$, which is about 1/100 of the Galactic mass.
Figure \ref{line} gives the merger rate of BHs along the Galactic plane. Due to the density profile of DM (See Eq. \ref{dm}),
the closer is the binary BHs to the Galactic center, the more efficiently do the binary BHs  accrete DM.
Based on the orbital decay of gravitational release (See Eq. \ref{gr}), the larger is binary BHs, the more rapid is gravitational release.
Therefore, the merger rate have an weak increase near the Galactic center. This increase is lower than 10\%.
We estimate that the range of the merger rate is about $55\rm Gpc^{-3}yr^{-1}$ ($Z=0.02$)
to $197 \rm Gpc^{-3}yr^{-1}$ ($Z=0.001$), which is consistent with the estimate of \cite{Abbott2016a}.
Comparing Figure-\ref{fig2} and Figure-\ref{line}, one can find that the enhancement of the mass at mergers and the enhancement of the mergers rates all occur at $r\sim0.1 \rm pc$. The dark matter accretion has a bigger influence on the mass of binary BHs than on the mergers rates.

{\begin{figure*}
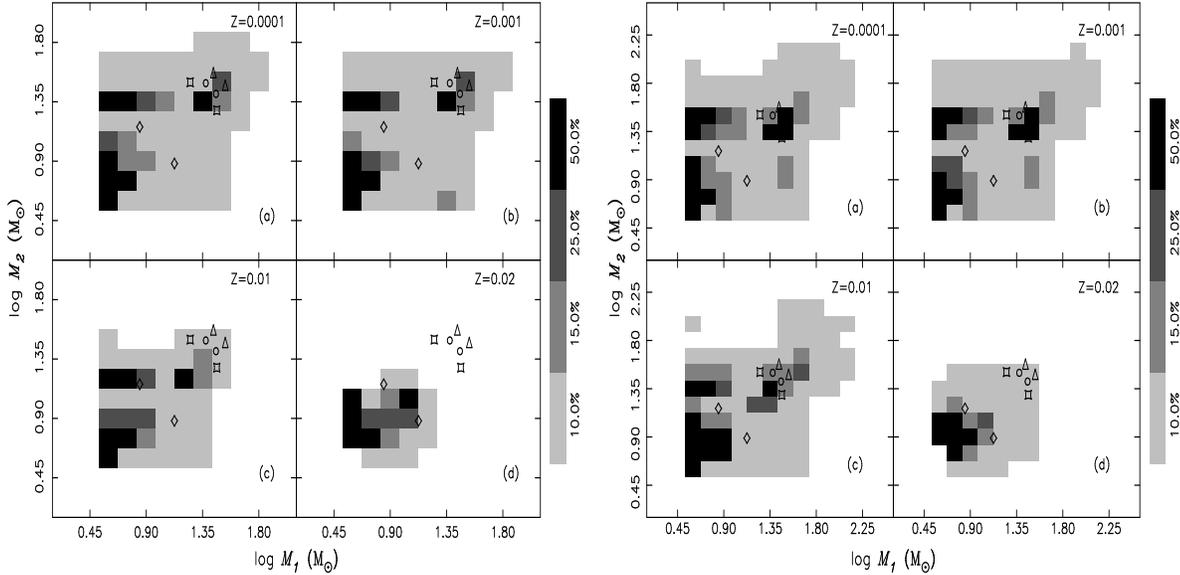

\begin{tabular}{lr}
\includegraphics[totalheight=3.0in,width=3.0in,angle=-90]{Figure-5.ps}&
\includegraphics[totalheight=3.0in,width=3.0in,angle=-90]{Figure-6.ps}\\
\end{tabular}
\caption{The binary BHs mass distribution without (left panel) and with (right panel) DM accretion.
The diamonds, boxes, circles and triangles represent the masses of binary BHs
in GW151226, GW170104, GW170814, and GW150914, respectively.
}
\label{fig:mass}
\end{figure*}
Figure \ref{fig:mass} shows the mass distributions of binary BHs  without and with the DM accretion
for different $Z$s. For the model without the DM accretion,
the results for the models with low metallicities ($Z=0.0001$ and 0.001) can
cover all observational data, while these for higher metallicities ($Z=0.01$ and 0.02)
can not explain the massive binary BHs. The main reasons are that the chemical homogeneous evolution
is more efficient for the rotating stars with low metallicity \citep{Heger2000,Mandel2016,Marchant2016}.
For the model with the DM accretion, the masses of binary BHs increase.
For the results of the models with $Z=0.0001$, 0.001, and 0.01 can cover all observations well.
However, the model with high $Z$ can not still explain most of the observations.

\section{Conclusion}
In this paper, we have investigated the merger rate of binary BHs
by constructing the model of DM accretion to binary BHs.
We estimate that the range of the merger rate is about $55 \rm Gpc^{-3}yr^{-1}$ to $197 \rm Gpc^{-3}yr^{-1}$.
The DM accretion have a very weak effect on the merger rate. However,
it can efficiently increase the masses of binary BHs.
In our model, the result for $Z = 0.01$ without the DM accretion can not explain GW170104, GW170814, and GW150914,
while with DM accretion, it can cover all observations well.
For the higher metallicity ($Z = 0.02$),
our model can not explain the mergers of high-mass binary BHs like GW170104, GW170814, and GW150914.

\section*{ACKNOWLEDGEMENTS}
We acknowledge the anonymous referee for careful reading of the paper and constructive criticism.
This work was supported by the National Natural Science Foundation
of China under Nos. 11473024, 11363005, 11763007, 11503008, 11365022,
and the XinJiang Science
Fund for Distinguished Young Scholars under No. QN2016YX0049.
{\color{red}{\software{MESA \citep{Paxton2011,Paxton2013,Paxton2015}, BSE \citep{Hurley2002}.}}}
\end{document}